\documentclass[aps,pra,twocolumn,showpacs,superscriptaddress]{revtex4-2}

\usepackage{graphicx}
\usepackage{amssymb}
\usepackage{amsmath}
\usepackage{color,xcolor}
\usepackage{bm}
\usepackage{physics}
\usepackage{ulem}

\usepackage{changes}

\begin{document}

\title{Complex energy structures of exceptional point pairs in two level systems}

\author{Jung-Wan Ryu}
\affiliation{Center for Theoretical Physics of Complex Systems, Institute for Basic Science (IBS), Daejeon 34126, Republic of Korea.}
\author{Chang-Hwan Yi}
\email{yichanghwan@hanmail.net}
\affiliation{Center for Theoretical Physics of Complex Systems, Institute for Basic Science (IBS), Daejeon 34126, Republic of Korea.}
\author{Jae-Ho Han}
\email{jaehohan@kaist.ac.kr}
\affiliation{Department of Physics, Korea Advanced Institute of Science and Technology (KAIST), Daejeon 34141, Republic of Korea.}

\begin{abstract}
We investigate the topological properties of multiple exceptional points in non-Hermitian two-level systems, emphasizing vorticity as a topological invariant arising from complex energy structures. We categorize EP pairs as fundamental building blocks of larger EP assemblies, distinguishing two types: type-I pairs with opposite vorticities and type-II pairs with identical vorticities. By analyzing the branch cut formation in a two-dimensional parameter space, we reveal the distinct topological features of each EP pair type. Furthermore, we extend our analysis to configurations with multiple EPs, demonstrating the cumulative vorticity and topological implications. To illustrate these theoretical structures, we model complex energy bands within a two-dimensional photonic crystal composed of lossy materials, identifying various EP pairs and their branch cuts. These findings contribute to the understanding of topological characteristics in non-Hermitian systems.
\end{abstract}

\maketitle

\section{Introduction}

Exceptional points (EPs) are a distinctive feature of non-Hermitian systems, where both eigenvalues and eigenstates coalesce \cite{Kato1976perturbation, Heiss1990avoided, Lee2008divergent, Ryu2009coupled, Dietz2011exceptional, Ding2015coalescence, Cui2019exceptional}. These points represent critical topological phenomena and have attracted considerable interest due to their unique properties and potential applications in fields such as optics, photonics, acoustics, and open quantum systems \cite{Xu2016topological, Shi2016accessing, Miao2016orbital, Ding2016emergence, Chen2017exceptional, Yang2018antiferromagnetism, Oezdemir2019Parity, Miri2019exceptional}. Adiabatic deformation of a system along a path encircling one or more exceptional points (EPs) in parameter space induces a nontrivial exchange among the corresponding eigenstates. This eigenstate switching behavior allows for classifying EPs based on the conjugacy class of the permutation and braid group \cite{Wang2021topological, Hu2021knots, Li2021quantized, Hu2022knot, Wojcik2022eigenvalue, Patil2022measuring, Ryu2022classification, Zhang2023observation}, a process known as homotopy classification.

Unraveling the behavior of EPs and their topological characteristics is essential to advancing our understanding of non-Hermitian systems. Specifically, the study of EP pairs has been a central focus in both theoretical models and experimental implementations. While early theoretical studies examined EP pairs with the same vorticities and no shared branch cuts \cite{Heiss2012the} (see Appendix), most observed EP pairs have opposite vorticities with shared branch cuts between them \cite{Zhou2018observation, Su2021direct, Ferrier2022unveiling, Zhang2022universal, Park2022double, Ryu2024exceptional, Kozii2024non, Ryu2024realization}. The key topological distinction between different types of EP pairs lies in their vorticity, a crucial topological invariant related to the braiding of complex eigenenergies. Vorticity describes how eigenenergies wind around EPs in the complex plane, playing a fundamental role in characterizing these singular points. Recently, numerous theoretical and experimental studies have explored the complex distributions of a large number of EPs \cite{Luitz2019exceptional,Zhong2023numerical,Erb2024novel}.

In this work, we reveal two distinct global and local structures of complex eigenenergies in parameter space, leading to corresponding global and local vorticities of EPs. We investigate the behavior of EP pairs with opposite vorticities (type-I) and those with the same vorticities (type-II) in two-dimensional parameter spaces, focusing on the formation of branch cuts and the role of vorticity as a topological invariant of EPs. Additionally, we demonstrate the realization of various types of EP pairs in complex energy bands using a photonic crystal composed of lossy materials and discuss the results.

\section{Topological structures of complex eigenenergies associated with EPs}
\label{sec:theory}

\subsection{Vorticity of an EP}
\label{sec:oneEP}

\begin{figure}
\centering
\includegraphics[width=\linewidth]{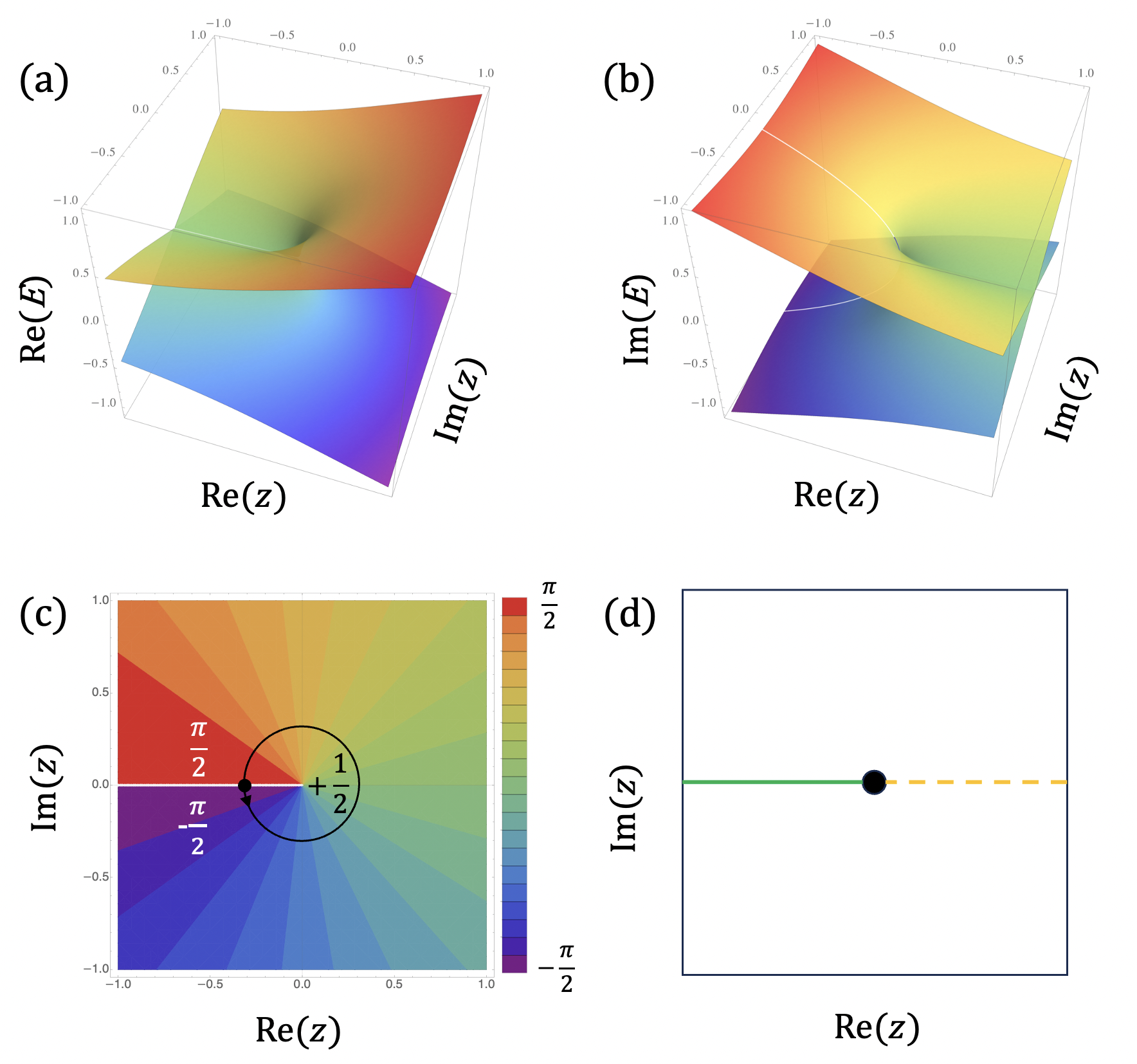}
\caption{(a) Real parts and (b) imaginary parts of complex energy $E_{\pm}$ of Eq.~(\ref{eq:oneEP}) when $z_1 = 0$. (c) Arguments of complex energy difference $E_{+}-E_{-}$. The argument increases by $\pi$ after a one-cycle counterclockwise journey on an encircling loop (black circle) since the white line in the arguments represents $\pm \pi /2$. The vorticity of an EP at $z=0$ is $+1/2$. (d) Schematic diagram of an EP (black circle) and real (green straight line) and imaginary (orange dashed line) branch cuts on 2D parameter space.}
\label{fig1}
\end{figure}

Two complex eigenenergies and their corresponding eigenstates coalesce at an EP, around which the eigenenergies $E_{\pm}$ exhibit a square root branch behavior, 
\begin{equation}
E_{\pm} = \pm \sqrt{z - z_1} ,
\label{eq:oneEP}
\end{equation}
where the complex parameter $z$ represents two-dimensional coordinates (Re($z$), Im($z$)) and $z_1$ is the parameter set of the EP in non-Hermitian systems [Fig.\ref{fig1}]. In the two-dimensional parameter space, the EP gives rise to corresponding real and imaginary branch cuts [Fig.\ref{fig1}(d)], which are topologically protected in the two-dimensional space \cite{Ryu2024pseudo}. If we consider a closed loop $\Gamma$ encircling the EP, the vorticity $\nu_1(\Gamma)$ can be defined as follows \cite{Leykam2017edge, Shen2018vortex}: 
\begin{equation}
\nu_{1} (\Gamma) = - \frac{1}{2 \pi} \oint_{\Gamma} \nabla_{\mathbf{k}} \mathrm{arg}[E_{+} (\mathbf{k}) - E_{-} (\mathbf{k})] \cdot d \mathbf{k} . 
\label{eq:vorticity}
\end{equation}
Here, $\nu_1$ corresponds to the winding number of the eigenenergies $E_{+}$ and $E_{-}$ around the EP in the complex-energy plane. By substituting $z - z_1 = r e^{i \theta}$ and $E_{\pm} = \pm \sqrt{r} e^{i \theta / 2}$, we obtain $\nu_1 = \pm 1/2$ for a closed loop $\Gamma$ encircling an EP. The $\pm$ signs of $\nu_1$ indicate counterclockwise and clockwise rotations on the complex-energy plane, respectively. In non-Hermitian systems, the vorticity is a topological invariant associated with the eigenenergy structures, such as the Berry phases from eigenstates, since the vorticity does not depend on the paths of closed loops unless EPs are added or removed inside the loop. 
\subsection{EP pairs}
\label{sec:twoEPs}

\begin{figure}
\includegraphics[width=\linewidth]{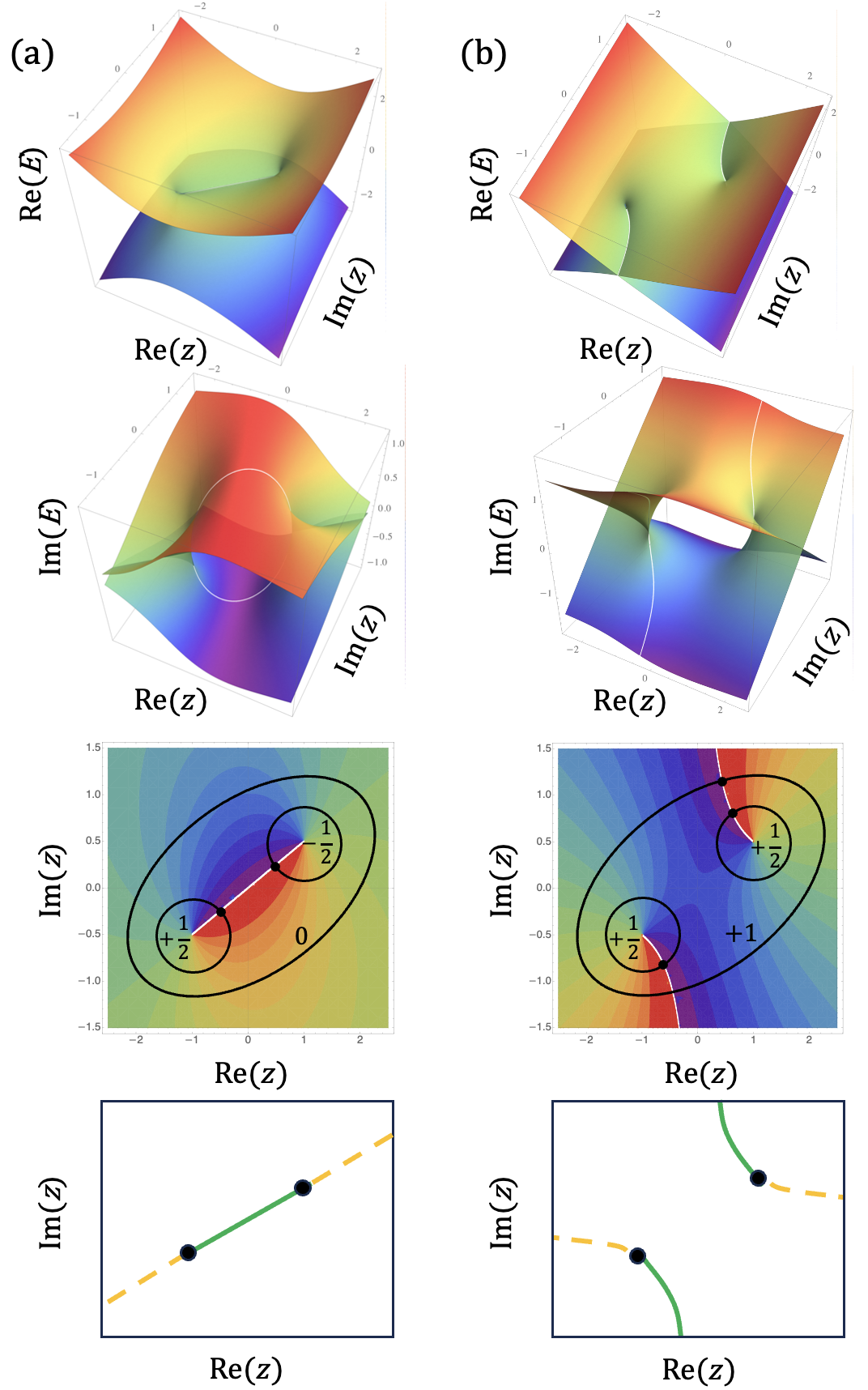}
\caption{(a) Real parts and imaginary parts of complex energy $E_{\pm}$ of Eq.~(\ref{eq:twoEP_1}), arguments of complex energy difference $E_{+}-E_{-}$, and schematic diagram of EPs (black circle) and real (green lines) and imaginary (orange dashed lines) branch cuts on 2D parameter space when $z_1 = -1.0 - 0.5 i$ and $z_2 = 1.0 + 0.5 i$. The color scale of arguments is the same as Fig.~\ref{fig1}(c). The vorticity of each EP are $\pm 1/2$ and the total vorticity of the two EPs is $0$. (b) Real parts and imaginary parts of complex energy  $E_{\pm}$ of Eq.~(\ref{eq:twoEP_2}), arguments of complex energy difference $E_{+}-E_{-}$, and schematic diagram when $z_1 = -1.0 - 0.5 i$ and $z_2 = 1.0 + 0.5 i$. The vorticity of each EP are $1/2$ and the total vorticity of the two EPs is $1$.}
\label{fig2}
\end{figure}

Next, we consider a closed loop that encircles two EPs. There are two types of EP pairs: counter-rotating EPs, which have opposite vorticities, and co-rotating EPs, which have the same vorticities. These are referred to as type-I and type-II EP pairs, respectively. The eigenenergy structures for type-I and type-II EP pairs are topologically equivalent to the energy structures given by
\begin{eqnarray}
\label{eq:twoEP_1}
E_{\pm}^{\rm I} &=& \pm \sqrt{( z - z_1 )( z - z_2 )^{*}} ~~~~ \mathrm{and}  \\
E_{\pm}^{\rm II} &=& \pm \sqrt{( z - z_1 )( z - z_2 )} ,
\label{eq:twoEP_2}
\end{eqnarray}
where $z_1$ and $z_2$ are the two EPs.

For the type-I pair, since $\nu_1 = 1/2$ for $\sqrt{( z - z_1 )}$ and $\nu_2 = -1/2$ for $\sqrt{( z - z_2 )^{*}}$, the vorticity of the loop $\Gamma$ encircling the two EPs results in zero after one complete cycle [Fig.~\ref{fig2}(a)]. If the two EPs move closer and eventually merge, i.e., $z_1 \rightarrow z_0$ and $z_2 \rightarrow z_0$, the two EPs coalesce into a Dirac point (DP) with eigenenergies $E_{\pm}^{\rm I} = \pm \left| z - z_0 \right|$. A Dirac point can appear in the two-dimensional parameter space since symmetry is restored when the two EPs merge. A shared real (or imaginary) branch cut exists between the two EPs, while two independent imaginary (or real) branch cuts originate from each EP. The topological structure of the closed loop $\Gamma$ remains unchanged even when the two EPs merge and form a DP, which has zero vorticity. For example, the eigenenergies of non-Hermitian Su–Schrieffer–Heeger (SSH) model can be rewritten as Eq.~(\ref{eq:twoEP_1}) around the EP pairs (see Appendix). 

For the type-II pair, since $\nu_{1,2} = 1/2$ for both $\sqrt{( z - z_1 )}$ and $\sqrt{( z - z_2 )}$, the loop $\Gamma$ encircling the two EPs results in a vorticity of $\nu = 1$ after one cycle [Fig.~\ref{fig2}(b)]. When the two EPs merge, they form a vortex point (VP) with eigenenergies $E_{\pm}^{\rm II} = z - z_0$, where $\nu = 1$ \cite{Shen2018vortex}. The vortex points cannot be removed on two-dimensional spaces by continuous transformation and symmetry breakings such as EPs and branch cuts \cite{Ryu2024pseudo}, unlike Dirac points lifted by symmetry breakings. There are no shared branch cuts between the two EPs, and each EP independently maintains its real and imaginary branch cuts. As exceptional cases, there seems to be shared branch cut when two branch cuts cross each other (see Appendix). In this work, the cases where $E_{\pm} = \pm \sqrt{( z - z_1 )^{*}( z - z_2 )}$ and $E_{\pm} = \pm \sqrt{( z - z_1 )^{*}( z - z_2 )^{*}}$ correspond to type-I and type-II pairs, respectively, as the sign of the vorticity depends on the chosen convention.

The total vorticity of multiple EPs within a closed loop is the sum of the individual vorticities of each EP enclosed by the loop. In multi-level systems, the sum rule of vorticities is applied to EPs that share the same energy levels. The EPs with different energy levels are independent. For example, if EP${_1}$ and EP${_2}$ correspond to the first and second states, and the second and third states, respectively, in a three-level system, the vorticities of the two EPs are independent: the vorticity of EP$_2$ can be $\pm 1/2$ irrespective to the vorticity of EP$_1$. Thus, we concentrate on the two-state systems in this study.

\subsection{EP Triples}
\label{sec:threeEPs}

\begin{figure}
\includegraphics[width=\linewidth]{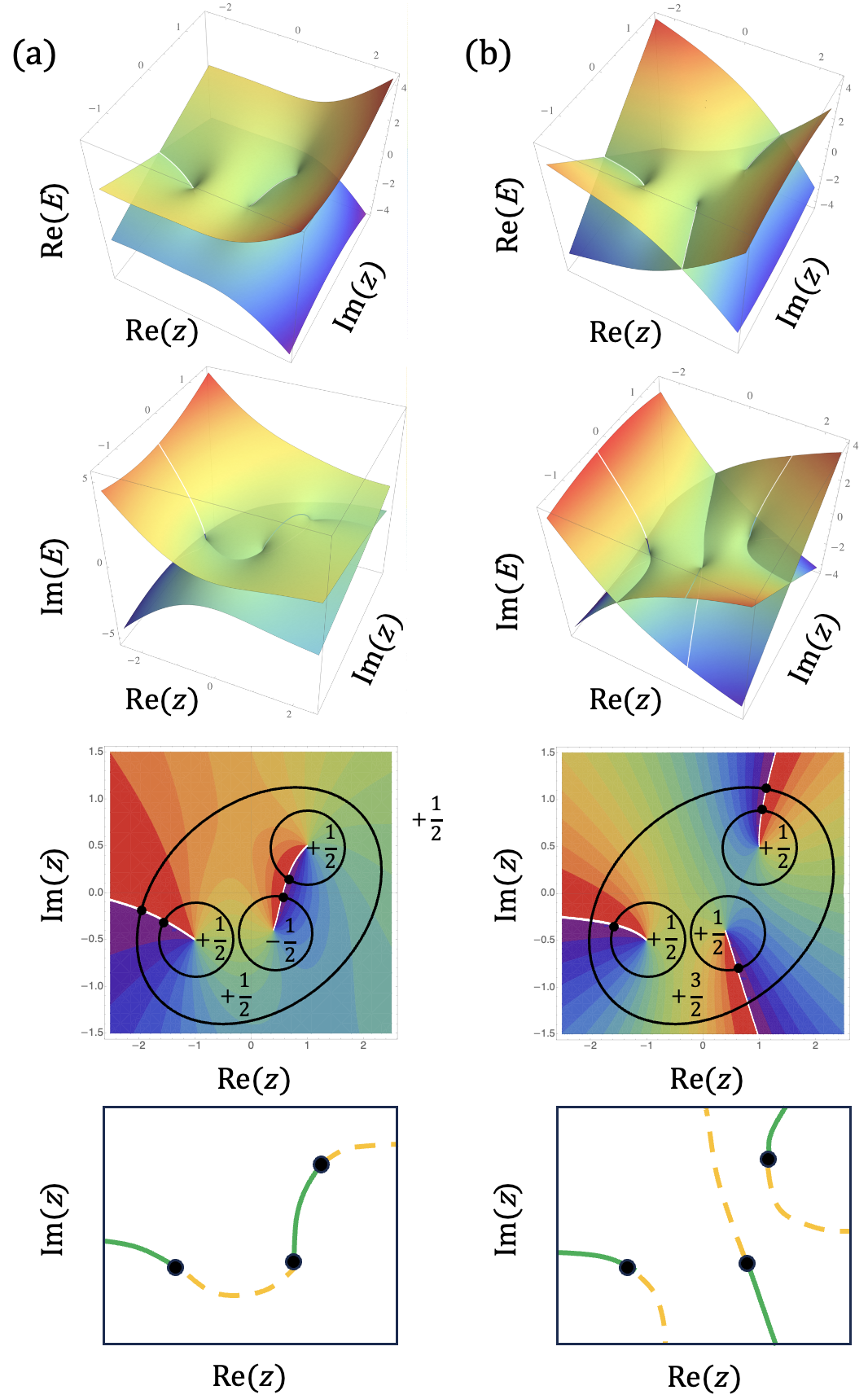}
\caption{(a) Real parts and imaginary parts of complex energy $E_{\pm}$ of Eq.~(\ref{eq:threeEP_21}), arguments of complex energy difference $E_{+}-E_{-}$, and schematic diagram of EPs (black circle) and real (green lines) and imaginary (orange dashed lines) branch cuts on 2D parameter space when $z_1 = -1.0 - 0.5 i$, $z_2 = 1.0 + 0.5 i$, and $z_3 = 0.4 - 0.4 i$. The color scale of arguments is the same as Fig.~\ref{fig1}(c). The vorticity of each EP are $\pm 1/2$ and the total vorticity of the three EPs is $1/2$. (b) Real parts and imaginary parts of complex energy  $E_{\pm}$ of Eq.~(\ref{eq:threeEP_30}), arguments of complex energy difference $E_{+}-E_{-}$, and schematic diagram when $z_1 = -1.0 - 0.5 i$, $z_2 = 1.0 + 0.5 i$, and $z_3 = 0.4 - 0.4 i$. The vorticity of each EP are $1/2$ and the total vorticity of the three EPs is $3/2$.}
\label{fig3}
\end{figure}

Considering a closed loop encircling three EPs, there are two possible configurations. The first configuration consists of two co-rotating pairs and one counter-rotating pair of EPs, while the second consists of three co-rotating EPs [Figs.~\ref{fig3}(a) and (b)]. These configurations are referred to as type-\{2,1\} and type-\{3,0\} triples, respectively. Here, type-$\{n,m\}$ indicates the energy of $n$ number of holomorphic factors and $m$ number of anti-holomorphic factors in the square root. The eigenenergies of type-\{2,1\} and type-\{3,0\} triples can be expressed as: \begin{eqnarray}
\label{eq:threeEP_21}
E_{\pm}^{\{2,1\}} &=& \pm \sqrt{( z - z_1 )( z - z_2 )( z - z_3 )^{*}} ~~~~ \mathrm{and}  \\
E_{\pm}^{\{3,0\}} &=& \pm \sqrt{( z - z_1 )( z - z_2 )( z - z_3 )} ,
\label{eq:threeEP_30}
\end{eqnarray}
where $z_1$, $z_2$, and $z_3$ represent the positions of the three EPs.

For the type-\{2,1\} triple, $\nu_{1,2} = 1/2$ for both $\sqrt{( z - z_1 )}$ and $\sqrt{( z - z_2 )}$, while $\nu_3 = -1/2$ for $\sqrt{( z - z_3 )^{*}}$. As a result, the vorticity of the loop $\Gamma$ encircling all three EPs is $1/2$ after a complete cycle [Fig.\ref{fig3}(a)]. When a loop encircles only two EPs, the resulting vorticity and branch cut formation correspond to either the type-I or type-II pairs. EP pairs, rather than individual EPs, serve as robust building blocks for constructing the topological structures of multiple EPs. This is because the EPs and their associated branch cuts are topologically protected in the two-dimensional parameter space, and what matters is not the absolute value of the vorticity, but whether the vorticities of the two EPs are the same or opposite, indicating co-rotating or counter-rotating behavior. For the type-\{3,0\} triple, since $\nu_{1,2,3} = 1/2$ for all three EPs $\sqrt{( z - z_1 )}$, $\sqrt{( z - z_2 )}$, and $\sqrt{( z - z_3 )}$, the loop $\Gamma$ encircling the three EPs results in a vorticity of $3/2$ after a complete cycle [Fig.~\ref{fig3}(b)].

\subsection{EP multiples}
\label{sec:nEPs}

Finally, we remark on the $n$ EPs cases, where their eigenenergies are given by
\begin{equation}
\label{eq:nEPs}
E_{\pm} = \pm \sqrt{( z - z_1 )^{(*)}( z - z_2 )^{(*)}( z - z_3 )^{(*)} \cdots ( z - z_n )^{(*)}} .
\end{equation}
The vorticity clearly reflects the parity of the number of EPs. For even $n$, the number of possible configurations of $n$ EPs with $\pm 1/2$ vorticities is $(n+2)/2$, and the global vorticity, which is the total sum of the vorticities of all EPs, is an integer. For example, in the case of two EPs, the global vorticity is either $0$ or $1$. For four EPs, the global vorticity is $0$ for two co-rotating and two counter-rotating EPs, $\pm 1$ for three co-rotating and one counter-rotating EP, and $\pm 2$ for four co-rotating EPs.

For odd $n$, the number of possible configurations of $n$ EPs is $(n+1)/2$, and the global vorticity, which is the total sum of the vorticities of all involved EPs, is a half-integer. For instance, the global vorticity is $\pm 1/2$ or $\pm 3/2$ in the case of three EPs. The half-integer vorticity represents the defectiveness of degeneracy in non-Hermitian two-level systems, while the integer vorticity is not related to the defectiveness of the degeneracy.

\subsection{Branch cuts on encircling loops}
\label{sec:BCs}

Different types of EP pairs and EP triples can be understood through the products of both branch cut operations, $\mathcal{O}_R$ and $\mathcal{O}_I$, which represent state exchanges crossing real and imaginary branch cuts, respectively, along encircling loops. These operations are directly related to the braiding of complex eigenenergies. To study the braiding of states, it is convenient to label the states by the state indices of their real and imaginary energy components. We use the notation ($i$, $j$) to denote the state associated with the $i$-th real energy and the $j$-th imaginary energy, where the energies are ordered by descending values. For instance, the state index with the largest real and imaginary energies is labeled as ($1$, $1$). When closed loops are considered, intersections occur between these loops and the branch cuts connected to EPs. As parameters move along these loops and cross real or imaginary branch cuts, the state index ($1$, $1$) transitions to ($2$, $1$) or ($1$, $2$), respectively.

\begin{figure}
\includegraphics[width=\linewidth]{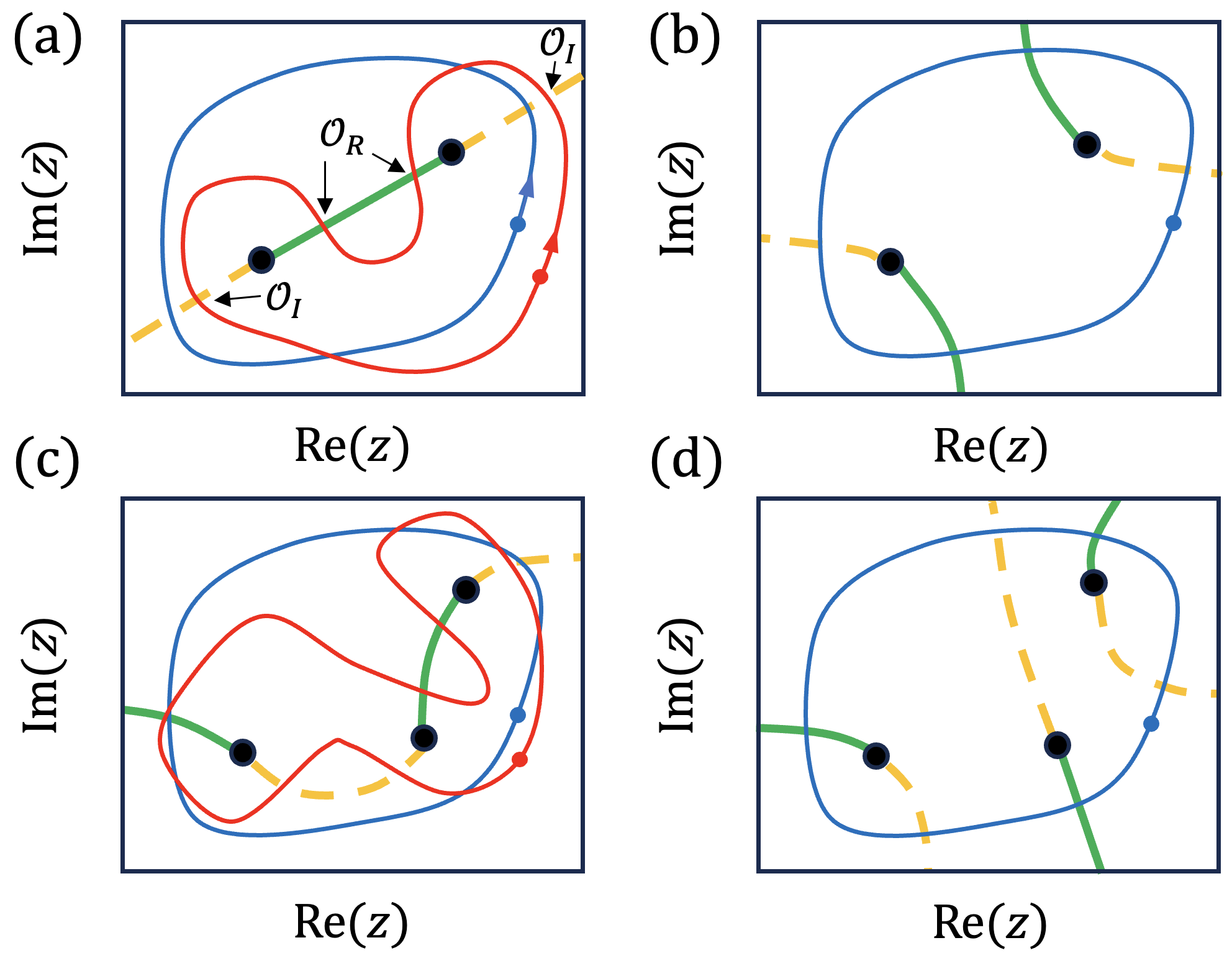}
\caption{Exceptional points (black circles), branch cuts (green lines and orange dashed lines), and selected encircling loops (blue and red closed loops) in the cases of (a) type-I and (b) type-II EP pairs and (c) type-\{2,1\} and (d) type-\{3,0\} EP triples. The parameters change on the encircling loops from the initial parameters (blue and red circles) counterclockwise. The loop crosses real and imaginary branch cuts are described by operations, $\mathcal{O}_R$ and $\mathcal{O}_I$, respectively.}
\label{fig4}
\end{figure}

Now, we demonstrate that the topology of the complex energy structure can be effectively represented by the string of operators $\mathcal O_R$ and $\mathcal O_I$. It is important to note that the combinations of these strings are unique for any loops encircling the same EPs. For the type-I EP pairs shown in Fig.~\ref{fig4}(a), the (1,1) state transitions to (1,2) when the blue loop crosses an imaginary branch cut, and then returns to (1,1) when it crosses a second imaginary branch cut. These parameter changes along the blue loop can be represented as the product of branch cut operations, $\mathcal{O}_I \mathcal{O}_I = I$, where $I$ is the identity operation. The red loop yields the operation $\mathcal{O}_I \mathcal{O}_R \mathcal{O}_R \mathcal{O}_I = I$, since $\mathcal{O}_R \mathcal{O}_R = \mathcal{O}_I \mathcal{O}_I = I$.

For type-II EP pairs, as shown in Fig.~\ref{fig4}(b), the (1,1) state changes to (1,2) at an imaginary branch cut, to (2,2) at the next real branch cut, to (2,1) at the following imaginary branch cut, and finally back to (1,1) at the last real branch cut. Although the initial and final states are the same as in the type-I case, the branch cut operation, $\mathcal{O}_R \mathcal{O}_I \mathcal{O}_R \mathcal{O}_I$, differs from that of type-I pairs. This operation corresponds to a non-zero vorticity, $\nu = \pm 1$, since $\mathcal{O}_R \mathcal{O}_I$ for an EP implies $\nu = \pm 1/2$. Therefore, the operations for EP pairs are $I$ for type-I (corresponding to $\nu = 0$) and $\mathcal{O}_R \mathcal{O}_I \mathcal{O}_R \mathcal{O}_I$ for type-II (corresponding to $\nu = \pm 1$).

For type-\{2,1\} EP triples in Fig.~\ref{fig4}(c), the branch cut operations are $\mathcal{O}_R \mathcal{O}_I$ for the blue loop and $\mathcal{O}_I \mathcal{O}_I \mathcal{O}_R \mathcal{O}_R \mathcal{O}_R \mathcal{O}_I$ for the red loop, both yielding identical results. In the case of type-\{3,0\} EP triples shown in Fig.~\ref{fig4}(d), the resulting operation is $\mathcal{O}_R \mathcal{O}_I \mathcal{O}_R \mathcal{O}_I \mathcal{O}_R \mathcal{O}_I$. Therefore, there are two distinct operations for EP triples: $\mathcal{O}_R \mathcal{O}_I$ for type-\{2,1\} (corresponding to $\nu = \pm 1/2$) and $\mathcal{O}_R \mathcal{O}_I \mathcal{O}_R \mathcal{O}_I \mathcal{O}_R \mathcal{O}_I$ for type-\{3,0\} (corresponding to $\nu = \pm 3/2$).

Operations involving both real and imaginary branch cuts describe the braiding of complex eigenenergies, while those involving only one type of branch cut, either real or imaginary, reduced to the permutation description \cite{Cartarius2009exceptional, Ryu2012analysis, Zhong2018winding}. For example, type-I and type-II EP pairs are represented by $I$ and $\mathcal{O}_R \mathcal{O}_I \mathcal{O}_R \mathcal{O}_I$, respectively. If only the real parts of the complex eigenenergies are considered, the operation for type-II EP pairs simplifies to $\mathcal{O}_R \mathcal{O}_R = I$, making type-I and type-II EP pairs identical in terms of permutation.

\section{Photonic crystal with lossy material}
\label{sec:realization}

\begin{figure}
\includegraphics[width=\linewidth]{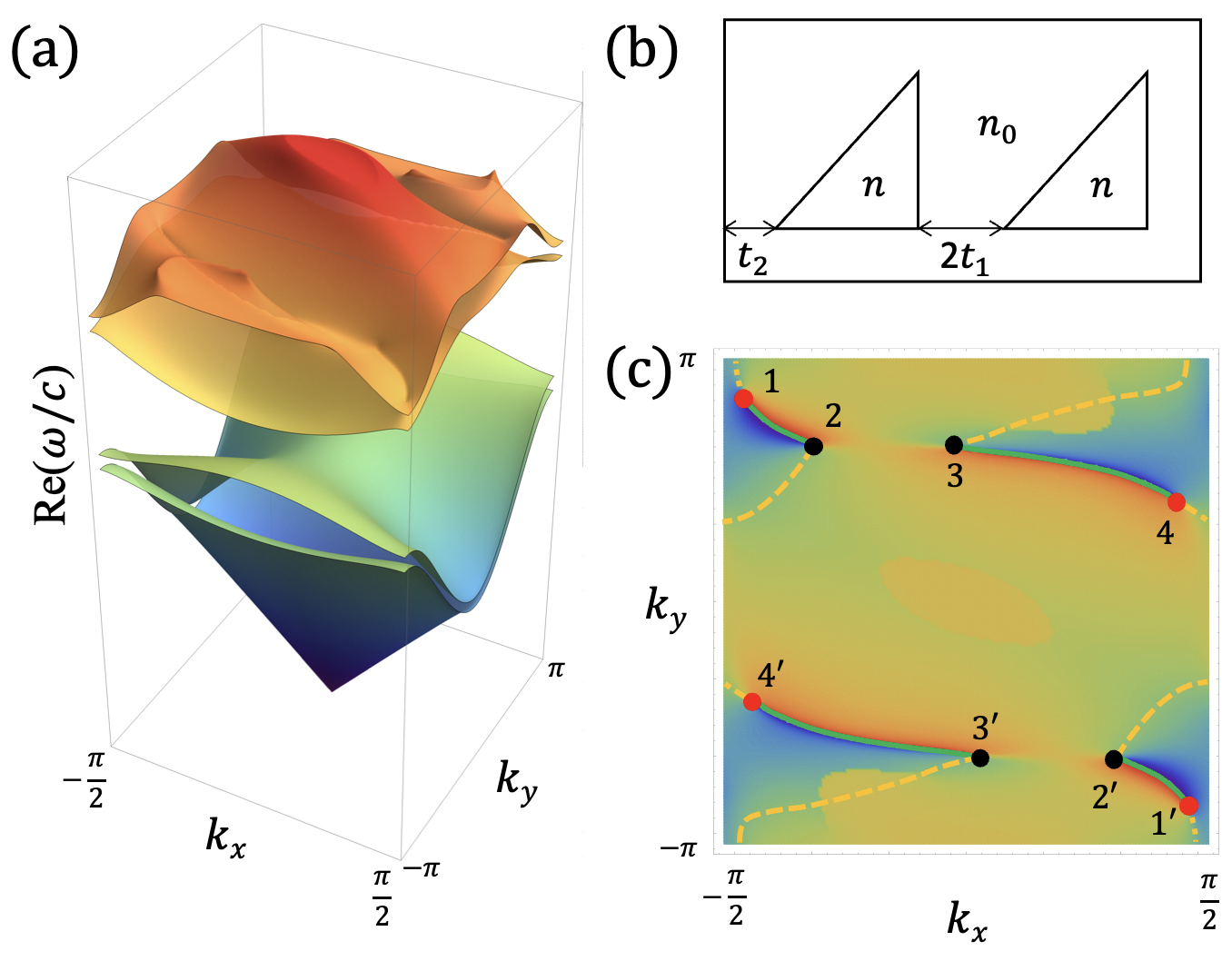}
\caption{(a) The lowest four energy bands in (b) a photonic crystal made of two lossy triangular cavities with a complex refractive index $n$. (c) The arguments of complex energy difference between third and fourth energy bands. The color scale of arguments is the same as Fig.~\ref{fig1}(c). It is noted that the discontinuity of arguments results from those of imaginary parts of third and fourth energy bands, defined by the order of real parts of energies. The black and red circles represent EPs with positive and negative vorticities, respectively. The green straight and orange dashed lines do simplified real and imaginary branch cuts, respectively.}
\label{fig5}
\end{figure}

To realize various pairs of EPs, we consider the complex energy bands in a 2D photonic crystal composed of two lossy cavities. The cavity losses introduce effective non-reciprocity, enabling the study of nontrivial point-gap topology, the non-Hermitian skin effect, and multiple EPs \cite{Zhong2021nontrivial, Zhong2023numerical}. We consider a lattice of triangular cavities with two bases in each unit cell, as shown in Fig.~\ref{fig5}(b). The positions of the two triangular cavities are represented by  $t_1=t_0-\delta$ and $t_2=t_0+\delta$, where the asymmetry parameter is $\delta=0.05$ (in units of the lattice constant). The complex refractive index of the cavities is $n=2-0.8i$, where the imaginary part represents the loss. Maxwell's equations for this system are solved using the finite-element method.

The real eigenvalues of the lowest four bands are shown in Fig.~\ref{fig5}(a), where we focus on the third and fourth bands. There are eight EPs associated with these two bands, connected by branch cuts [see Fig.~\ref{fig5}(c)]. The eight EPs (EP1, EP2, EP4, EP3, EP1$'$, EP2$'$, EP4$'$, and EP3$'$) are alternately connected by real and imaginary branch cuts due to the periodicity of the two-dimensional Brillouin zone. Neighboring EP pairs along this chain, consisting of EPs and branch cuts, are type-I pairs with opposite vorticities. As $\delta$ increases, the EP pairs (EP1-EP2, EP1$'$-EP2$'$, EP3-EP4, and EP3$'$-EP4$'$) move closer and eventually meet. At critical values of $\delta$, they merge into Dirac points ($\nu = 0$), and the bandgap opens as $\delta$ further increases. Conversely, as $\delta$ decreases, EP1-EP1$'$, EP2-EP3, EP2$'$-EP3$'$, and EP4-EP4$'$ move closer and transition into vortex points ($\nu = \pm 1$) when $\delta = 0$. For $\delta \neq 0$, these pairs are type-II pairs with the same vorticities.

\section{Summary}
\label{sec:summary}

We examined the topological structures of complex eigenenergies associated with pairs of exceptional points in non-Hermitian two-level systems, identifying two distinct types: type-I pairs with opposite vorticities and type-II pairs with identical vorticities. These EP pairs exhibit unique topological properties, particularly regarding the presence or absence of shared branch cuts. Extending this framework, we analyzed the topology and cumulative vorticity in configurations with three or more EPs, demonstrating that the total vorticity within an encircling loop is determined by the sum of individual EP vorticities. Also, we show that the complex energy structure and braiding of state exchanges can be represented by a string of consecutive exchange operators on real and imaginary branch cuts, $\mathcal O_{R}$ and $\mathcal O_I$. To support these theoretical insights, we modeled complex energy band structures within a photonic crystal composed of lossy materials, observing multiple EP pairs and their branch cuts. At critical parameter values, these EP pairs converge into Dirac or vortex points. This study provides valuable insights into the topological invariants of EPs in non-Hermitian systems, offering a deeper understanding of their impact on complex energy band structures.

\section*{acknowledgments}
We acknowledge financial support from the Institute for Basic Science in the Republic of Korea through the project IBS-R024-D1.

\setcounter{figure}{0}
\setcounter{equation}{0}

\renewcommand\thefigure{A\arabic{figure}}
\renewcommand\theequation{A\arabic{equation}}

\section*{Appendix}

\subsection{An example of type-II EP pair}
\label{review_EPs}

We briefly review the two-dimensional matrix model for level repulsion where there are a pair of EPs \cite{Heiss2012the}. Consider the problem
\begin{eqnarray}
H (z) &=& H_0 + H_1 (z) = H_0 + z V \\\nonumber
      &=& 
\begin{pmatrix}
\omega_{+} & 0 \\
0 & \omega_{-}
\end{pmatrix}
+ z
\begin{pmatrix}
\epsilon_{+} & \delta \\
\delta & \epsilon_{-}
\end{pmatrix}
\end{eqnarray}
where the parameters $\omega_{\pm}$ and $\epsilon_{\pm}$ determine the non-interacting resonance energies $E_{\pm} = \omega_{\pm} + z \epsilon_{\pm}$. Owing to the interaction invoked by the matrix element $\delta$ the two levels do not cross but repel each other. In fact, the levels turn out to be
\begin{eqnarray}
E_{\pm} (z) &=& \frac{1}{2} \Big(\omega_{+} + \omega_{-} + z (\epsilon_{+}+\epsilon_{-}) \\\nonumber 
&\pm& \sqrt{(4 \delta^2 + (\epsilon_{+} - \epsilon_{-} )^2) (z-z_{1})(z-z_{2})} \Big)
\end{eqnarray}
and the two levels coalesce for complex values of z in the vicinity of the level repulsion, that is at
\begin{eqnarray}
    z_1 = \frac{i (\omega_{+} - \omega_{-} )}{-2 \delta -i (\epsilon_{+} - \epsilon_{-} )} \\
    z_2 = \frac{i (\omega_{+} - \omega_{-} )}{2 \delta -i (\epsilon_{+} - \epsilon_{-} )} ,
\end{eqnarray}
which are two EPs. The $E_{\pm} (z)$ has a form $\sqrt{ (z - z_1) (z - z_2) }$ and their energy structures are topologically the same as those of type-II EP pairs in Fig.~\ref{fig2} (b), denoting two EPs are co-rotating. As the $\delta$ approaches zero, two EPs become closer. If $\delta = 0$, $E_{\pm} (z)$ has a form $z - z_1$, which does not exhibit the square-root singularity.

\begin{figure}
\includegraphics[width=\linewidth]{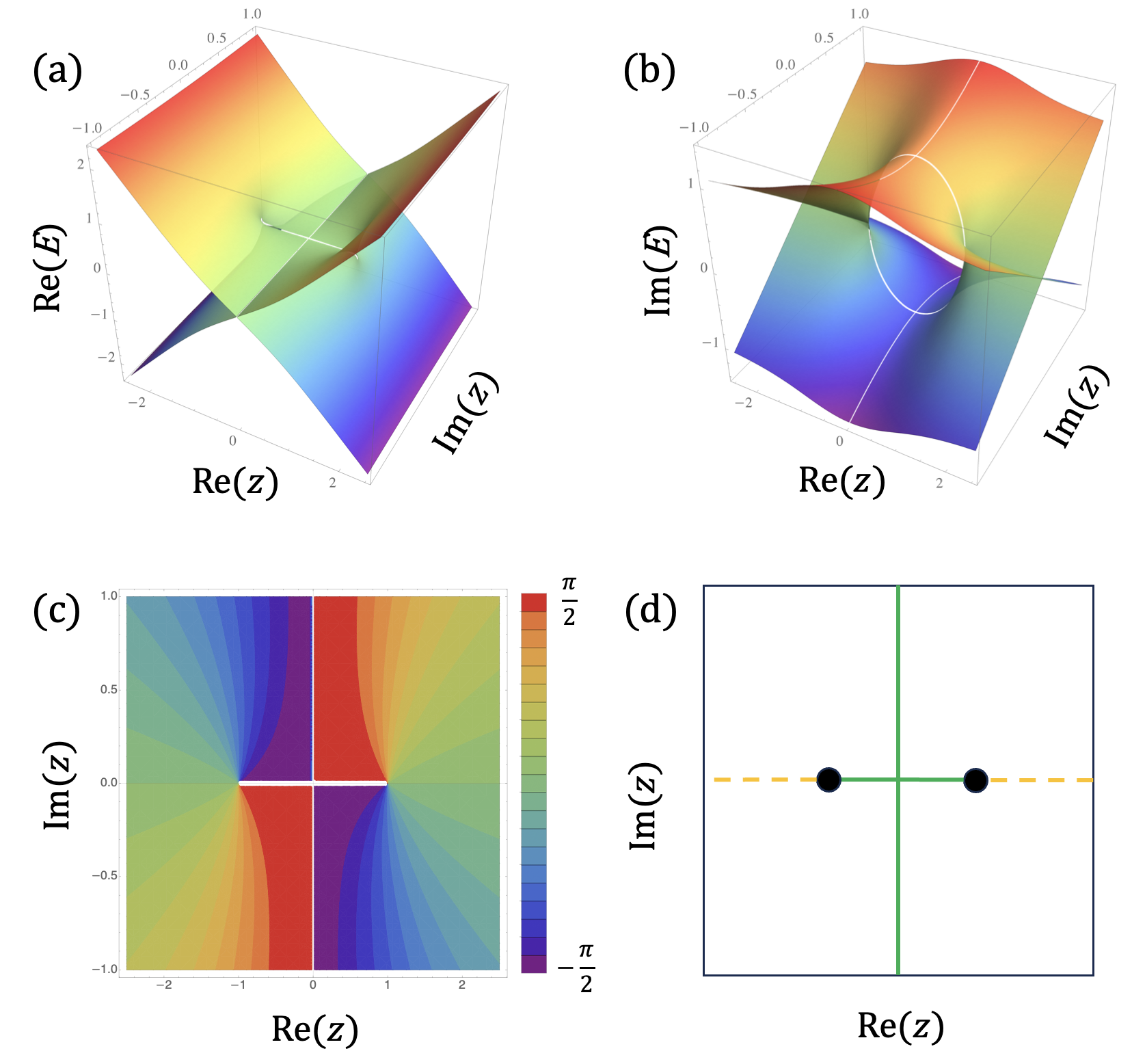}
\caption{(a) Real parts and (b) imaginary parts of complex energy $E_{\pm}$ of Eq.~(\ref{eq:twoEP_2}) when $z_1 = -1$ and $z_2 = 1$. (c) Arguments of complex energy difference $E_{+}-E_{-}$. The color scale of arguments is the same as Fig.\ref{fig1}(c). The vorticity of each EP are
$1/2$ and the total vorticity of the two EPs is $1$. (d) Schematic diagram of two EPs (black circle) and real (green straight lines) and imaginary (orange dashed line) branch cuts on 2D parameter space.}
\label{figa1}
\end{figure}

\subsection{EP pairs in non-Hermitian Su--Schrieffer--Heeger model}
\label{NHSSH}

We consider the non-Hermitian Su–Schrieffer–Heeger (SSH) model, where the Hamiltonian is given as \cite{Ryu2024exceptional}
\begin{equation}
    h(k) = 
\begin{pmatrix} 
 0 & d + t e^{\theta} e^{-i k} \\
 d + t e^{-\theta} e^{i k} & 0 \\
\end{pmatrix}\ .
\label{eq_3by3}
\end{equation}
Here, $d$ represents intra-unit-cell hopping, $t$ denotes inter-unit-cell hopping, and $\theta$ describes the non-reciprocity between left and right directional hopping. The eigenenergies are
\begin{equation}
    E_{\pm} = \pm \sqrt{(e^{\theta} + t e^{i k})(e^{-\theta} + t e^{-i k})} .
\end{equation}
There are two EPs at $(t,k) = (e^{\theta}, \pi)$ and $(e^{-\theta}, \pi)$. The energies can be rewritten as
\begin{equation}
    E_{\pm} = \pm \sqrt{(z - z_1 )(z - z_2 )^{*}},
\end{equation}
where $z = - t e^{i k} + i \pi$, $z_1 = e^{\theta} + i \pi$, and $z_2 = e^{-\theta} + i \pi$. Their energy structures are topologically the same as those of type-I EP pairs in Fig.~\ref{fig2} (a), denoting two EPs are counter-rotating. As the $\theta$ approaches zero, two EPs become closer. If $\theta = 0$, $E_{\pm}$ has a form $\abs{z - z_1}$, which corresponds to the Dirac point.

\subsection{Exceptional case of type-II EP pairs}
\label{ExCase}

Figure~\ref{figa1} shows the exceptional case that there seems to be shared branch cuts between two EPs with the same vorticities. The branch cut crossing can be realized when real or imaginary parts of $z_1$ and $z_2$ in Eq.~(\ref{eq:twoEP_2}) are the same, while branch cuts repel each other generally as shown in Fig.~\ref{fig2}(b).

\end{document}